\documentstyle[aps,manuscript]{revtex}
\begin{document}
\draft
\preprint{}

\title{First-Order Vortex Lattice Melting and Magnetization of
YBa$_2$Cu$_3$O$_{7-\delta}$}
\author{R. \v{S}\'{a}\v{s}ik and D. Stroud}
\address{
Department of Physics,
The Ohio State University, Columbus, Ohio 43210}

\date{\today}

\maketitle

\begin{abstract}

We present the first non-mean-field calculation of the magnetization $M(T)$
of YBa$_2$Cu$_3$O$_{7-\delta}$ both above and below the flux-lattice melting
temperature $T_m(H)$. The results are in good agreement with experiment
as a function of transverse applied field $H$.
The effects of fluctuations in both order parameter $\psi({\bf r})$
and magnetic induction $B$ are included
in the Ginzburg-Landau free energy
functional:
$\psi({\bf r})$ fluctuates within the lowest Landau level in each layer,
while $B$ fluctuates uniformly according to the appropriate
Boltzmann factor. The second derivative $(\partial^2 M/\partial T^2)_H$ is
predicted to be
negative throughout the vortex liquid state and positive in the solid state.
The discontinuities in entropy and magnetization
at melting are calculated to be
$\sim 0.034\, k_B$ per flux line per layer and
$\sim 0.0014$~emu~cm$^{-3}$ at a field of 50 kOe.

\end{abstract}


\newpage

Even more than perfect conductivity, the magnetization ${\bf M}$
is a sensitive probe of the superconducting state.
Type-I superconductors have an abrupt jump in
magnetic susceptibility from 0 to $-1/4\pi$
at the superconducting transition.  Low-$T_c$ type-II superconductors
have a more gradual behavior:
the magnetization sets in continuously at the upper critical field
$H_{c2}(T)$, reaching $-H/4\pi$ only at the lower critical field
$H_{c1}(T)$.  In high-$T_c$ superconductors, ${\bf M}$ also
turns on continuously, but deviates from mean-field predictions. ${\bf M}$
may also be affected by the flux-lattice melting
transition\cite{blatter,farrell0} which occurs
well below the mean-field superconducting transition temperature.
To predict these deviations from mean-field behavior
is a stringent test of any theory of fluctuations in the superconducting
state.

In this paper we calculate the magnetization of the most studied high-$T_c$
material, YBa$_2$Cu$_3$O$_{7-\delta}$, including the effects of fluctuations.
To our knowledge, this is the first non-mean-field calculation of ${\bf M}$
in YBa$_2$Cu$_3$O$_{7-\delta}$.
We consider an applied field ${\bf H}$ parallel to the $c$ axis, and
also calculate the first-order flux-lattice melting curve $T_m(H)$
for the same material.  Both ${\bf M}$
and $T_m(H)$ are in very good agreement with experiment over a range
of magnetic fields.

Our approach is to start from a Ginzburg-Landau free energy functional which
includes the field energy in the form\cite{tinkham}
\begin{equation}
G[\psi,{\bf A}] = \int\, d^3{\bf r} \left\{
a(T,z)|\psi({\bf r})|^2 +
\frac{1}{2m^*}\left|\left(-i\hbar \nabla-\frac{e^*}{c}{\bf A}({\bf r})
\right)\psi({\bf r})\right|^2
+\frac{b}{2}|\psi({\bf r})|^4 + \frac{({\bf B}-{\bf H})^2}{8 \pi}
\right\}.
\end{equation}
Here $\psi({\bf r})$ is the complex order parameter,
${\bf A}({\bf r})$ the vector potential, ${\bf B} = {\bf \nabla} \times
{\bf A}$, $e^* = 2e$ is the charge of a Cooper pair,
${\bf H} \equiv H\hat{z}$ is the applied magnetic field intensity, and
$a(T,z)$, $b$, and $m^*$ are phenomenological parameters.
The periodic $z$\/-dependence of $a(T,z)$ is introduced to
characterize the underlying layered structure.
We assume that spatial modulation of $a(T,z)$, which acts as
a potential for Cooper pairs, results in a tight-binding form for
$\psi({\bf r})$
along the $z$\/-axis, thus creating the effective mass anisotropy $\gamma$.
The integral is to be carried out over a fixed sample volume
$V=(\phi_0/H)N_\phi N_z s$, where $s$ is the periodicity of
the layered structure, $\phi_0 = hc/2e$ is the flux quantum, $N_\phi$ is the
number of flux lines threading the sample, and $N_z$ is the number of layers.
The Gibbs free energy density
appropriate to an experiment at fixed $T$ and ${\bf H}$
is
\begin{equation}
{\cal G} = -(k_BT/V) \ln Tr_{\psi, {\bf A}} \exp( -G[\psi, {\bf A}]/k_BT ).
\end{equation}
We also define free energy `per flux line per layer' according to
${\cal G}_\phi \equiv (\phi_0 s/H){\cal G}$; internal energy $G_\phi$ and
entropy $S_\phi$ are defined in the same fashion.

To determine the magnetization, we
assume a {\em uniformly fluctuating}
magnetic induction $\nabla \times {\bf A}({\bf r}) \equiv B\hat{z}$, so that
$B$ can in principle assume different values
in the flux liquid and solid phases (but remains parallel to the $c$ axis).
This approximation is most reasonable in the extreme type-II limit ($\kappa
\gg 1$)
when density correlations among the vortices are well developed and vortices
are clearly defined, as expected in the neighborhood of the melting
line; it may deteriorate very near the upper critical field $H_{c2}(T)$.
In addition, the assumption of uniform ${\bf B}\| c$ should
be best in the vortex liquid phase, where contributions to the
local field ${\bf B}({\bf r})$ come from many positionally
uncorrelated vortex line
segments, as suggested by Brandt\cite{brandt}, but may still be reasonable
in the vortex solid phase, even though
${\bf B}$ in that phase should develop a nonzero variance
$\sigma^2 = \langle {\bf B}({\bf r})^2 \rangle-
\langle {\bf B}({\bf r}) \rangle^2$.

We evaluate the statistical average (2) using the following procedure.
First, we expand the order parameter $\psi({\bf r})$ in a basis
which consists of products of lowest Landau level states of the operator
$(2m^*)^{-1}(-i\hbar\nabla_\perp -e^*{\bf A}/c)^2$ in the $ab$ plane,
and Wannier functions from
the lowest band of states of the operator $a(T,z)-(2m^*)^{-1}\hbar^2
\partial^2/\partial z^2$
in the $c$ direction\cite{sasik}.  This leads to an explicit form for
all but the last term of the integrand in eq.\ (1), in terms of the
complex coefficients $c_{k,n}$ of the expansion, corresponding to the
$k^{th}$ lowest Landau state in the $n^{th}$ layer\cite{sasik}.
In order to impose periodic boundary conditions in the $ab$ plane, we use
the Landau gauge\cite{hu}. Then the statistical average
implied by eq.\ (2) is carried out
by a Monte Carlo (MC) procedure, in which the coefficients $c_{k,n}$,
{\em and also the average magnetic induction B},
are considered as fluctuating MC variables.
The entire procedure is closely analogous to the ``constant pressure''
MC ensemble well known in classical fluids\cite{mc}.  The variables analogous
to pressure and volume are $H$ and $B$.  The MC step
which changes $B$ increases or decreases the area
of the $ab$ plane at constant vortex number.

Within this approach, the {\em mean-field approximation}
to (2) is specified
by a pair $\psi({\bf r})$ and $B$, for which $G$ is minimum\cite{tinkham}.
In the normal state ($H>H_{c2}(T)$) this is achieved for $\psi({\bf r}) = 0$
and $B = H$. In the mixed state, ($H<H_{c2}(T)$), the corresponding minimum
is attained for a triangular lattice of straight vortex lines and magnetization
\begin{equation}
M \equiv \frac{B-H}{4\pi} =\frac{H-H_{c2}(T)}{4\pi(2\kappa^2{\beta_A} -1)},
\end{equation}
where ${\beta_A} = 1.15959\ldots$ is the Abrikosov ratio.  In the limit
$\kappa \gg 1$
studied here, this formula becomes identical to the original Abrikosov result,
which has $4\pi(2\kappa^2-1){\beta_A}$ in the denominator.  Thus our approach
and approximations are indeed correct at the mean-field level in this limit.
The corresponding mean-field free energy density is ${\cal G}^{MF}
= -(8\pi)^{-1}[H_{c2}(T)-H]^2/(2\kappa^2{\beta_A}-1)$.

In order to execute the MC calculation for YBa$_2$Cu$_3$O$_{7-\delta}$, we use
the following set of parameters: $T_{c0} = 93$~K,
$dH_{c2}(T)/dT = -1.8\times 10^4$~Oe/K, $s = 11.4$~\AA,
$\gamma = 5$, and $\kappa = 52$.  Although there is some experimental
evidence that $\kappa$ varies with magnetic field\cite{welp1}, we neglect
this field-dependence here.  In most of our calculations, we have considered
a cell containing $N_\phi = 10^2$ vortices and $N_z=10$ layers.
Our results are based typically on
2--3$\times10^5$ MC passes through the entire sample
following $\sim 2\times 10^4$ MC passes for equilibration.

Fig.\ 1 shows the calculated magnetization $M(T)$ of
YBa$_2$Cu$_3$O$_{7-\delta}$ as
a function of temperature $T$ at three different values of $H$.
The mean-field predictions are shown
for comparison.  There is no sign of a true phase transition at
the mean-field transition temperature $T_{c2}(H)$, since
$(\partial M/\partial T)_H$ is continuous at that point.  Instead, there is
an apparently first-order melting transition at a lower temperature
$T_m(H)$, as signaled by a weak discontinuity in the magnetization
curves (denoted by arrows in the Figure).  The melting curve $T_m(H)$
inferred from this discontinuity is shown in the
inset to Fig.\ 1; it is
in good agreement with experiment.

The general behavior of $M(H, T)$ agrees very well with
experiment\cite{welp2}.  For example, the calculated second derivative
$(\partial^2 M/\partial T^2)_H$ is negative throughout
the vortex liquid phase, in
accord with recent measurements based on a differential torque
technique\cite{farrell1}.
A second-degree polynomial fit to our magnetization results just above
the melting temperature $T_m$ yields
$(\partial^2 M/\partial T^2)_H = (-0.0038 \pm 0.002)$
{}~emu~cm$^{-3}$K$^{-2}$ for $H=20$~kOe, and
$(\partial^2 M/\partial T^2)_H = (-0.0030 \pm 0.002)$
{}~emu~cm$^{-3}$K$^{-2}$ for $H=50$~kOe.
Experiment\cite{farrell1} also gives a negative
$(\partial^2 M/\partial T^2)_H$
for fields in the range 10--20~kOe, and of the same order of magnitude.
In the solid phase, we find
our calculated $(\partial^2M/\partial T^2)_H >0$.

Another striking feature of our results is the crossing of the
magnetization curves as a function of temperature.
This crossing is also observed
in experiment at a similar temperature\cite{welp2}.  In
Bi$_2$Sr$_2$CaCu$_2$O$_{8+\delta}$, the same phenomenon
has attracted much theoretical attention\cite{bulaevskii,txbls}, because it
is believed to occur at a unique temperature $T^*$ independent of field.
In YBa$_2$Cu$_3$O$_{7-\delta}$,
experimental data of Welp {\em et al.}\cite{welp2} reveal
that magnetization curves do not cross at a single point, presumably
because, although layered, YBa$_2$Cu$_3$O$_{7-\delta}$
is only moderately anisotropic.
As may be seen in the Figure, our calculated magnetization curves
also fail to cross at a single point.   Furthermore,
the order of the calculated crossings as a function of field is the same as
observed in YBa$_2$Cu$_3$O$_{7-\delta}$\cite{welp2}.

The second derivative of the magnetization
can also be calculated using a Maxwell relation
\begin{equation}
\left( \frac{\partial^2 M}{\partial T^2}\right)_H = \left(
\frac{\partial (C_H/T)}{\partial H}\right)_T,
\end{equation}
where $C_H \equiv -T(\partial^2 {\cal G}/\partial T^2)_H$ is the specific heat
of the sample at constant $H$\cite{caveat}.
Since eq.\ (4) equates the second derivative of one macroscopic quantity
to the first derivative of another, it provides a potentially
more precise way of determining $(\partial^2M/\partial T^2)_H$ than a direct
calculation of $M(T)$.  Note that in the mean-field approximation
$C_H^{MF} \equiv -T(\partial^2 {\cal G}^{MF}/\partial T^2)_H =
(T/4\pi)(dH_{c2}/dT)^2/(2\kappa^2{\beta_A} -1)$, which is independent of
$H$. This
implies, via eq.\ (4), that $M^{MF}(T)$ is linear in $T$, in agreement with
eq.\ (3).
To go beyond mean-field theory, we have done an MC calculation of $C_H(H,T)$,
using the fluctuation-dissipation theorem\cite{fd}.
Our results are presented in Fig.\ 1 (inset).  Clearly, $d(C_H/T)/dH < 0$
in the vortex liquid phase, which implies, in agreement with
experiment\cite{farrell1}, that $M(T)$ has negative
curvature {\em throughout} the liquid phase, not just in the vicinity of
the mean-field critical temperature $T_{c2}(H)$.
The straight lines in the Figure
are constructed using slopes determined from independent calculations of
the magnetization.  To an excellent approximation, they are
tangent to the specific heat curves, as expected from eq.\ (4),
thus confirming the self-consistency of our approach.
Experimental specific heat data\cite{inderhees} confirm that
$d(C_H/T)/dH < 0$ in the vortex liquid phase.   In the solid phase, our work
predicts that $d(C_H/T)/dH > 0$.  This is in agreement
with recent, high-accuracy specific heat measurements by
Schilling and Jeandupeux\cite{schilling} on large twinned
YBa$_2$Cu$_3$O$_{7-\delta}$ crystals.
The interesting region in the immediate neighborhood of the melting line
has not been resolved, however.

Fig.\ 2 shows the in-plane structure factor $S({\bf q}_{\perp},0)$
defined as the thermal average
\begin{equation}
S({\bf q}_\perp, 0) = \left\langle \int d^3{\bf r}\, d^3{\bf r}'\,
|\psi({\bf r})|^2
|\psi({\bf r}')|^2 e^{i{\bf q}_\perp \cdot ({\bf r} - {\bf r}')}
\right\rangle,
\end{equation}
for a field $H = 50$ kOe at two temperatures $T = 82.8$~K and $T=83.0$~K,
corresponding to the vortex solid and vortex liquid phases.
This dramatic change occurs at the temperature of the magnetization
discontinuity, thus confirming that this discontinuity signals
a melting transition. In each case,
the displayed $S({\bf q}_{\perp},0)$ is the result of averaging over
100 configurations from the thermal ensemble.  For clarity, we have removed
the central maximum at ${\bf q}_{\perp}=0$, and have divided by the
``atomic'' structure factor $\exp(-q^2_{\perp}\ell^2/4)$, where
$\ell = (\phi_0/2\pi B)^{1/2}$ is the magnetic length.
The regular periodic structure of the maxima in Fig.\ 2(a) corresponds to an
ordered crystalline phase, while the concentric rings of Fig.\ 2(b)
characterize an isotropic fluid.  In both cases, the deviations from
perfect isotropy can be attributed to the finite size of the sample, as well
as to its rectangular shape.

Finally, we address the issue of the order of the melting transition.
As indicated by Fig.\ 2, the vortex solid state is a phase of
discrete symmetry, whereas the vortex liquid is an isotropic phase of much
higher symmetry. Therefore, upon melting, the vortex ensemble undergoes a
{\em discontinuous} symmetry change.
By the well-known argument of Landau\cite{landau}, the vortex
melting transition has to be first-order, with a finite jump in entropy
per unit volume $\Delta S$.  A critical point (defined by $\Delta S =0$)
cannot exist, and the liquid-solid phase boundary must terminate by
intersecting either the coordinate axes or other phase boundaries.

To calculate $\Delta S$, we use a variant of the histogram method
of Lee and Kosterlitz\cite{lee}.
In principle, one should resolve the energy distribution $P(G)$
into two Gaussian peaks at $T_m$, then confirm that the dip between the peaks
scales like a surface energy with increasing sample size.  Such a
calculation was done in Ref.\cite{hetzel} for the 3D XY model.  In the present
case, this would be a formidable computational effort, because
$\Delta S$ is small and the two peaks are not resolved at any accessible
system sizes.  Instead, we perform a long MC run
($\sim 10^6$ passes through the entire lattice)  near $T_m$, starting
from the system ground state.  During the simulation, the system flips
$\sim 2$--4 times between the two states in equilibrium.  In our
standard diffusive sampling algorithm, we can identify continuous (in
MC ``time'') sequences of representatives belonging to the same
homogeneous phase.  The two phases can be clearly distinguished
by their structure factors and mean internal energy.

In Fig.\ 3 we plot the probability distribution of internal energy, $P(G)$,
for these two phases in equilibrium at $T_m \sim 83$~K and $H=50$~kOe,
at three system sizes.  The low-energy peak always corresponds to the
ordered vortex solid phase.
Since the entropy change $\Delta S_\phi$ {\em decreases} with
system size, the value $\Delta S_\phi \sim 0.034\,k_B$ should be
considered an {\em upper bound} to $\Delta S_{\phi}$
in the thermodynamic limit. $\Delta S_{\phi}(H=20$~kOe) has a similar
size dependence and is $\sim 30$\% smaller than
$\Delta S_{\phi}(H=50$~kOe). This is the expected trend,
since $T_m$ is higher at $H=20$~kOe, and typical superconducting parameters,
such as the condensation energy $G^{MF}$, decrease with increasing temperature.

{}From our calculated $\Delta S$ at melting and the computed
slope $(dH/dT)_m$ of the melting curve from Fig.\ 1, we can estimate the
magnetization jump $\Delta M$ at melting via
the Clausius-Clapeyron relation $\Delta S/\Delta M = -(dH/dT)_m$.
Inserting the calculated values of
this slope and of $\Delta S_\phi$, we obtain
$\Delta M \sim 0.0014$~emu~cm$^{-3}$ at $H = 50$~kOe, and
$\Delta M \sim 0.0005$~emu~cm$^{-3}$ at $H = 20$~kOe.
These values are consistent with the directly calculated $\Delta M$
seen in Fig.\ 1.  In experiment, no finite magnetization jump has yet been
observed\cite{farrell1,zeldov}, although transport measurements on
untwinned
YBa$_2$Cu$_3$O$_{7-\delta}$ crystals are widely interpreted as evidence for a
first-order melting transition\cite{safar1,safar2}.
The null result of Farrell
{\em et al.}\cite{farrell1} for $\Delta M$
puts an upper bound $\Delta S_\phi < 0.003\, k_B$
at $H=20$~kOe, a value almost ten times smaller than predicted in the present
work.  However, as suggested by these authors themselves, the presence of
defects may have suppressed
the measured $\Delta M$ to some extent.  This discrepancy remains to be settled
by an experiment which attains perfect reversibility in both temperature and
field.

To summarize, we have presented the first non-mean-field calculation of
magnetization in YBa$_2$Cu$_3$O$_{7-\delta}$,
using a constant-${\bf H}$
Monte Carlo technique in conjunction with a lowest Landau level approximation.
Our results yield a magnetization in very good agreement with experiment
in both the flux liquid and flux lattice state, as well as a melting curve
very close to experiment.  Our results provide perhaps the most detailed
evidence to date that a
Ginzburg-Landau free energy functional, based on a complex scalar order
parameter, adequately describes the thermodynamic properties
of YBa$_2$Cu$_3$O$_{7-\delta}$ near the flux lattice melting curve.

We are grateful for valuable conversations with Professors D. E. Farrell,
T. R. Lemberger, and W. F. Saam.  This
work was supported by NSF through Grant DMR94-02131, the Midwest
Superconductivity Consortium through DOE Grant DE-FG02-90ER45427.
The calculations were performed on the local network of DEC Alpha stations.

\newpage

\begin{center}
FIGURE CAPTIONS
\end{center}

\begin{enumerate}

\item Calculated magnetization $
M(T)$ of YBa$_2$Cu$_3$O$_{7-\delta}$ at three different applied
fields, $H=10$~kOe, 20 kOe, and $50$~kOe.
Dashed lines represent the mean-field solution (3).
Solid lines are  spline curves connecting the calculated points. Arrows
denote melting temperatures $T_m(H)$, as determined by the calculated
discontinuity in $M(T)$. Estimated errors in $M(T)$ are much smaller than
the symbol sizes. $N_\phi \times N_z = 10^2 \times 10$.\\
Right inset: locus of the liquid-solid
phase boundary in the {\em H-T} plane, as determined by our calculations and as
measured by Farrell {\it et al.}\cite{farrell2} and
Safar {\it et al.}\cite{safar1}.\\
Left inset:
specific heat $C_H$ as a function of magnetic field, taken at two
temperatures, $T=83$~K and 87~K.
The dashed line represents the mean-field value $C_H^{MF}/T$.
Arrows indicate the approximate location of
the fields at melting $H_m(T)$. Straight lines have slopes $-0.0038$~emu
cm$^{-3}$~K$^{-2}$ at 87 K and $-0.0030$~emu cm$^{-3}$~K$^{-2}$ at 83~K
(see text). $N_\phi \times N_z = 6^2 \times 6$.

\item (a) In-plane structure factor $S({\bf q}_\perp, 0)$, divided by the
``atomic''
structure factor $\exp (-q^{2}_{\perp}\ell^2/4)$, taken in the solid phase
at $T=82.8$~K and $H=50$~kOe, and averaged over 100 configurations.
The central maximum at ${\bf q}_\perp =0$ has been removed for clarity.

(b) Same as (a), but in the liquid phase at $T=83.0$~K.

\item Probability distribution $P(G)$ of internal energy at the finite-size
melting point $T_m \sim 83$~K, $H = 50$~kOe,
and three different system sizes.  For each system size, the
two separate peaks represent the
distributions of energy within the solid and the liquid
{\em at the same T and H}.  Because of the weak dependence of $T_m$
on system size, we have used for the zero of $G_\phi$ a size-dependent energy.
Horizontal bars denote the calculated
latent heats per flux line per layer, in units of $k_BT$.

\end{enumerate}


\begin{thebibliography}{99}
\bibitem{caveat} In actuality $C_H$ is just the superconducting part
of the specific heat of a crystal.
\end{thebibliography}
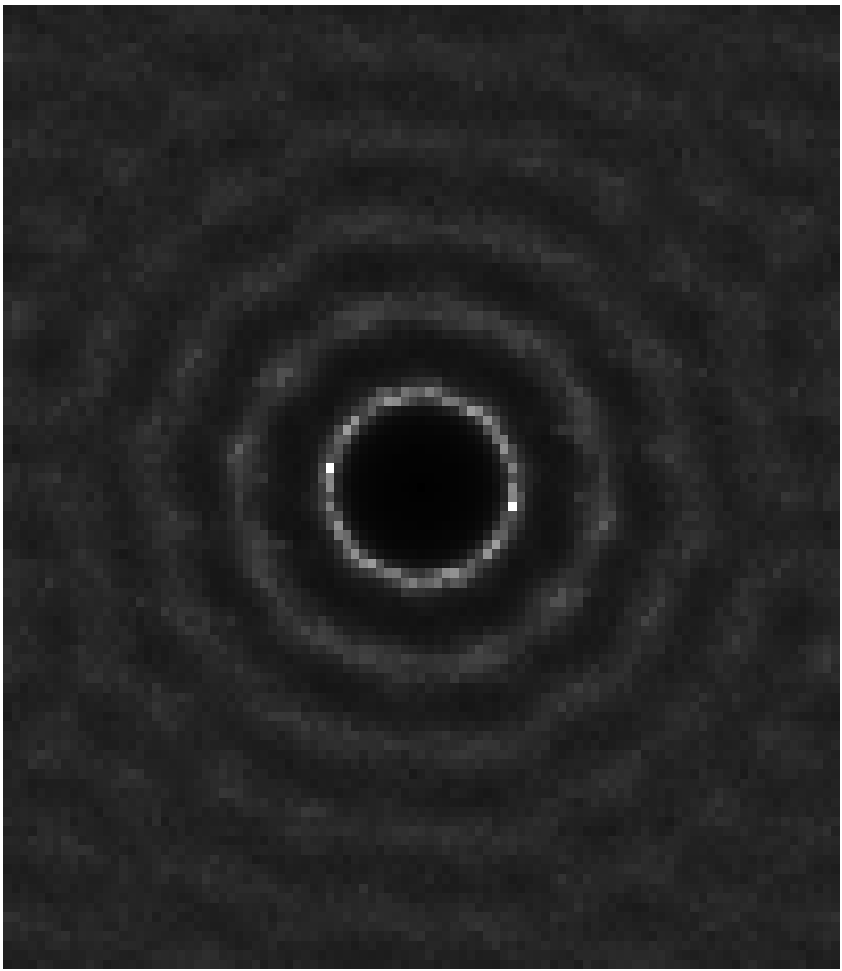
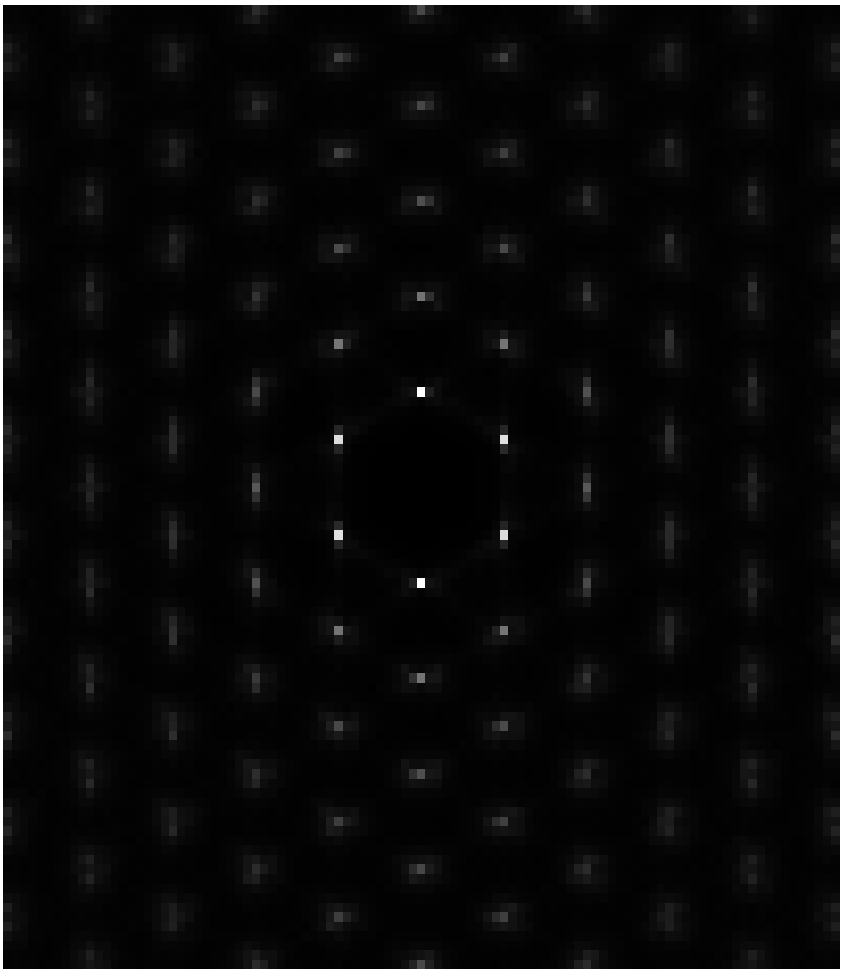
\end{document}